\documentclass[preprint]{aastex631}

\usepackage{color}
\usepackage{amsmath}

\begin{document}

\title{Constraints on Evolutions of Fundamental Constants from Clustering of Fast Radio Burst Dispersion Measure}

\correspondingauthor{Jun-Qing Xia}
\email{xiajq@bnu.edu.cn}

\author{Shi-Yuan Wang}
\affiliation{School of Physics and Astronomy, Beijing Normal University, Beijing 100875, China}
\affiliation{Institute for Frontiers in Astronomy and Astrophysics, Beijing Normal University, Beijing 100875, China}

\author{Jun-Qing Xia}
\affiliation{School of Physics and Astronomy, Beijing Normal University, Beijing 100875, China}
\affiliation{Institute for Frontiers in Astronomy and Astrophysics, Beijing Normal University, Beijing 100875, China}



\begin{abstract}
Constrained measurements of fundamental physical constants using astronomical observational data represent a powerful method for investigating potential new physics. In particular, the dispersion measure (DM) of fast radio bursts (FRBs), which probes the electron density along their propagation paths, may be influenced by the space-time variation of the fine-structure constant \(\alpha\). In this study, we analyze the cross-correlation signal between foreground galaxies and the DM of background FRBs to constrain the evolution of \(\alpha\). Assuming large-scale structure (LSS) galaxy surveys with the capabilities of the China Space Station Telescope (CSST) at \(z=0.15\) and { a mock FRB survey with \(N_{\text{FRB}}=10^5\) at \(z=0.4\), we test how well \(\alpha\) variation can be constrained}, with a standard deviation of \(\sigma(\Delta \alpha / \alpha) = 0.0007\) at \(z=0.15\).
Furthermore, taking into account the nonminimal coupling between the scalar field and the electromagnetic field, the variation in \(\alpha\) can lead to the non-conservation of photon number along geodesics. This would result in a violation of the CDDR and affect the evolution of the Cosmic Microwave Background (CMB) temperature. In this work, we { obtain constraints results} on the CDDR parameter \(\eta\) and the parameter \(\beta\) governing CMB temperature evolution at \(z=0.15\), yielding \(\sigma(\eta) = 0.0004\) and \(\sigma(\beta) = 0.0006\), respectively. Finally, we relate the variation in \(\alpha\) to the time evolution of the proton-to-electron mass ratio, { reporting a standard deviation} of \(\sigma(\Delta \mu/\mu) = 0.002\) at $z=0.15$. Future FRB surveys hold significant potential for advancing our understanding of the evolution of fundamental physical constants.
\end{abstract}

\keywords{Radio transient sources (2008), Cosmology (343)}


\section{Introduction} \label{sec:intro}

Dirac's Large Numbers Hypothesis posits that fundamental physical constants, such as \( e \), \( G \), \( \hbar \), \( m_e \), and \( m_p \), may vary over cosmological time, suggesting that the state of the universe could influence dimensionless cosmic constants \citep{1937Natur.139..323D}. This hypothesis contrasts with the prevailing assumption in most cosmological theories, which rely on the constancy of these fundamental parameters. Consequently, investigating the spatiotemporal evolution of these constants is of paramount importance in both physics and cosmology. Any deviation from the established theories would not only point to the presence of new physics \citep{2011LRR....14....2U,2003RvMP...75..403U} but also represent a significant departure from the \(\Lambda\)CDM model \citep{2017RPPh...80l6902M}, particularly in relation to phenomena such as the Hubble tension and \( S_8 \) tension \citep{2022MNRAS.510.2206H,2022JHEAp..34...49A}.

The fine-structure constant \(\alpha\), also referred to as the electromagnetic coupling constant, is a dimensionless quantity that appears in the equations governing the interaction between matter fields and electromagnetic fields. Any evolution of \(\alpha\) implies the non-conservation of photon number along geodesics \citep{2014PhRvD..90l4064H, 2013PhRvD..88b7506M}. Theoretically, if the coupling constants adhere to the principles of locality and general covariance, their dependence on spacetime is typically governed by a scalar field, which can subsequently lead to variations in physical constants \citep{Holanda:2015oda}. For instance, the runaway dilaton model, grounded in string theory, posits that the dilaton field \(\phi\), through its runaway behavior towards strong coupling, could induce variations in the fine-structure constant \citep{Martins:2015dqa, 2002PhRvD..66d6007D}. Directly measuring \(\alpha\) is crucial for testing dynamical dark energy models \citep{2017RPPh...80l6902M}, as it helps constrain the dynamics of the underlying scalar field. Moreover, \(\alpha\) can provide insights into the evolution of the universe at epochs when dark energy is not yet dominant. Additionally, studying the variation of \(\alpha\) can facilitate the investigation of cosmic distance duality relation (CDDR) violations, as well as CMB temperature evolution and distortion \citep{2014PhRvD..90l4064H}.

In recent years, significant research has been devoted to investigating the temporal variation of the fine-structure constant, particularly using astronomical data. For instance, observational data from Chandra X-ray and OVRO/BIMA interferometric Sunyaev-Zeldovich effect measurements have enabled \cite{Ferreira:2024kht} to constrain \(\Delta \alpha/\alpha\) with 1\% precision. The cosmic microwave background (CMB) has also been employed for such measurements \citep{Tohfa:2023zip, 2019PhRvD..99d3531S}. Specifically, \cite{2019PhRvD..99d3531S} applied the Bekenstein-Sandvik-Barrow-Magueijo (BSBM) model to CMB data from Planck 2018, achieving a constraint of 0.6\% precision. Furthermore, similar studies have been conducted using fast radio bursts (FRBs) \citep{2024arXiv240611691L} and quasar observations \citep{Wilczynska:2020rxx, LE2020e05011, 2024arXiv240901554W}.

A notable example within this context is provided by \cite{2024arXiv240611691L}, who utilized the dispersion measure (DM) of FRBs to constrain the evolution of the fine-structure constant. This study incorporated data from 17 well-localized FRBs, alongside simulated FRB data, and achieved a precision constraint as low as \(10^{-2}\). FRBs are a class of astrophysical phenomena, and their dispersion measure is a key indicator of the electron distribution along their propagation paths \citep{2023RvMP...95c5005Z, 2019A&ARv..27....4P, Lorimer:2007qn}. The measurement of DM suggests that FRBs likely originate from beyond the Milky Way, making them valuable tools for cosmological investigations.

In this letter, we assume that { a mock sample of FRBs} is located in a thin shell behind galaxies, with an effective redshift of \( z_f = 0.4 \), while the galaxy sample has an effective redshift of \( z_g = 0.15 \) and { a small redshift bin \( \Delta z_g = 0.3 \)}. We simulate the cross-correlation signal between the foreground galaxies and the dispersion measure (DM) of background FRBs, and perform a Markov Chain Monte Carlo (MCMC) analysis to constrain the evolution of the fine-structure constant. Section \ref{sec:2} introduces the relationship between the DM and the fine-structure constant, along with the power spectrum. In Section \ref{sec:3}, we discuss the impact of the evolution of \( \alpha \) on the CDDR, the CMB temperature evolution, and the time evolution of the proton-to-electron mass ratio. Finally, Section \ref{sec:4} presents the conclusions and discussions. In this work, we adopt the standard \( \Lambda \)CDM cosmological model \citep{2020A&A...641A...6P}.

\section{Methodology}\label{sec:2}

\subsection{Dispersion Measure of FRBs}

FRBs are electromagnetic waves, and as they travel through plasma, they experience dispersion. According to the dispersion relation, lower-frequency waves propagate more slowly than higher-frequency waves, causing a time delay in the observed signal on Earth. The DM quantifies the magnitude of this delay. A detailed derivation of DM can be found in \cite{2023RvMP...95c5005Z}.

First, the time for an electromagnetic wave with frequency $\omega$ to reach the observer at a redshift $z$ is given by:
\begin{equation}
	t_{\omega}=\int_0^{z^{\prime}}\frac{dz}{H(z)}\left(1+\frac12\frac{\omega_p^2}{\omega^2}\right)
\end{equation}
where $\omega_p\equiv\sqrt{{4\pi n_ee^2}/{m_e}}$ is the plasma frequency. Here, $n_e$ is the electron number density, $e$ is the elementary charge, and $m_e$ is the electron mass. From a cosmological perspective, when considering FRBs originating extra-galactically, we must account for the frequency shifts due to cosmological redshift \( z \), where the observed frequency \(\omega_{\text{obs}}\) is related to the emitted frequency \(\omega\): $\omega=(1+z)\omega_{\text{obs}}$. Thus, the arrival time difference between two observed frequencies, \( \omega_{\text{obs,2}} > \omega_{\text{obs,1} }\), can be expressed as:
\begin{equation}\label{eq:time delay}
	\Delta t = \frac{2\pi e^2}{m_ec}\left(\frac1{\omega_{\text{obs,1}}^2}-\frac1{\omega_{\text{obs,2}}^2}\right)\int_0^{z^{\prime}}\frac{dz}{H(z)}\frac{n_z(z)c}{(1+z)^2}~.
\end{equation}
The time delay between the two frequencies arises from the dispersion of the radio waves in the ionized intergalactic medium, and the observed time difference is inversely proportional to the square of the frequency. 

Then we can define the integral of free electrons number density along the line of sight, which consists of the contributions from the intergalactic medium as the ${\rm DM_{IGM}}$:
\begin{equation}\label{dm_igm}
	{\rm DM_{IGM}}=\int_0^{z^{\prime}}\frac{dz}{H(z)}\frac{cn_e(z)}{(1+z)^2}~.
\end{equation}
However, for a well-localized FRB, besides the ${\rm DM_{IGM}}$, the total DM also includes contributions from the Milky Way, \( \text{DM}_{\text{MW}} \), which can be determined through Galactic pulsar observations; and from the FRB host galaxy, \( \text{DM}_{\text{Host}} \), which is more challenging to quantify due to the dependence on the type of host galaxy, the relative orientation of the galaxy, and the local plasma environment, all of which are poorly understood.

Fortunately, in this study, we focus on the cross-correlation signal between the DM of background FRBs and foreground galaxies situated in a thin redshift shell. In this context, only \( \text{DM}_{\text{IGM}} \) contributes to a non-zero cross-correlation signal. The contributions from \( \text{DM}_{\text{MW}} \) and \( \text{DM}_{\text{Host}} \) are effectively nullified in this analysis, as these components are not aligned with the redshift distribution of the foreground galaxies. Therefore, in the following analysis we only consider the cross-correlation between \( \text{DM}_{\text{IGM}} \) and galaxies to constrain the evolution of fundamental constants like the fine-structure constant.

\subsection{Variation of Fine-structure Constant}

The fine-structure constant, denoted as \( \alpha \), is a dimensionless physical constant that characterizes the strength of the electromagnetic interaction. It reflects the intensity of the interaction between electrons and photons. The fine-structure constant is defined as:
\begin{equation}
\alpha = \frac{2 \pi e^2}{hc}~,
\end{equation}
where \( e \) is the elementary charge, \( \epsilon_0 \) is the permittivity of free space, { \( h \) is the Planck constant}, and \( c \) is the speed of light in vacuum.

In cosmological studies, it is often assumed that the fine-structure constant may vary over time, depending on the dynamics of the universe. The time evolution of \( \alpha \) can be expressed as:
{
\begin{equation}\label{eq:alpha1}
\frac{\Delta \alpha}{\alpha}(z) \equiv \frac{\alpha(z) - \alpha_0}{\alpha_0}~,
\end{equation}
where \( \alpha(z)=\alpha _0 +\Delta \alpha(z) \) is the value of the fine-structure constant at a given redshift \( z \), and \( \alpha_0 \) is the present-day value of the fine-structure constant.}

The variation in the fine-structure constant can arise from several theoretical frameworks, including scalar fields coupled to the electromagnetic field or from new physics beyond the Standard Model. Such variations would influence the time delay of FRBs, modifying the expression for the time delay in Eq. (\ref{eq:time delay}) as follows:
\begin{equation}
\Delta t = \frac{h \alpha _0}{m_e} \left( \frac{1}{\omega_{\text{obs,1}}^2} - \frac{1}{\omega_{\text{obs,2}}^2} \right) \int_0^{z'} \frac{dz}{H(z)} \frac{c n_e(z)}{(1+z)^2} \left( \frac{\Delta \alpha}{\alpha}(z) + 1 \right)~.
\end{equation}

As a result, the DM of FRBs, as described in Eq. (\ref{dm_igm}), will also be modified by multiplying the integral term by a factor of \( 1 + {\Delta \alpha}/{\alpha}(z) \).

\subsection{Power spectra of FRB and galaxy}

The DM along a given line of sight is expected to be correlated with the density of foreground galaxies in that direction, as some of the electron fluctuations contributing to the DM arise from these galaxies. Accordingly, our goal is to perform a cross-correlation between the foreground galaxy distribution and a map of DMs, as measured by FRBs.

For simplicity, we assume that the foreground galaxies reside within a thin redshift shell centered at $z_g = 0.15$ with { a shell width of $\Delta z_g = 0.3$}. The number density of these galaxies is $n_g = 3.4 \times 10^{-2} (\text{Mpc}/\text{h})^{-3}$, which is expected to be provided by the China Space Station Telescope (CSST) spectroscopic survey \citep{2019ApJ...883..203G}. { All mock FRBs}, acting as backlights for the free electrons in these galaxies, are assumed to lie within a thin redshift shell centered at $z_f = 0.4$ with a width of $\Delta z_f = 0.2$. Thus, the foreground galaxies are located within the redshift bin $(z_g - \Delta z_g/2, z_g + \Delta z_g/2)$, while the background FRBs are located within the redshift bin $(z_f - \Delta z_f/2, z_f + \Delta z_f/2)$. { In this thin-shell approximation, we treat the foreground galaxies and background DM fields as two-dimensional. The sufficiently large redshift separation between FRB host galaxies and foreground galaxies ensures that: (1) the DM is independent of the distribution of free electrons within the FRB host galaxy, (2) free electrons between the host and foreground galaxies do not affect the DM, and (3) there is no spatial correlation between FRB host galaxies and foreground galaxies. Therefore, any observed correlations between galaxy distributions and FRB DMs can be attributed to the interaction between the foreground galaxies and the free electron distribution along the line of sight.}

In the Limber approximation, we can calculate the auto-correlation angular power spectra of foreground galaxies $C_{\ell}^{\rm gg}$ and DMs of background FRBs $C_{\ell}^{\rm DD}$ and their cross-correlation angular power spectrum $C_{\ell}^{\rm Dg}$, which will be modified by the variation of fine-structure constant:

\begin{equation}
\begin{aligned}
	C_{\ell}^{\text{Dg}} &=  n_{\text{e0}}\frac{(1+z_g)}{\chi_{\text{g}}^{2}} \left(\frac{\Delta \alpha}{\alpha}(z_g)+1\right) P_{\text{ge}}\left(\frac{\ell+0.5}{ \chi_{g}}, z_{\text{g}}\right)\\
	C_{\ell}^{\text{DD}}  &= n^2_{\text{e0}}\int_0^{z_f} {d} z  \frac{(1+z)^2}{\chi^{2}(z)}\frac{c}{H(z)} \left(\frac{\Delta \alpha}{\alpha}(z)+1\right)^2P_{\text{ee}}\left(\frac{\ell+0.5}{ \chi_{g}}, z\right) \\
	C_{\ell}^{\text{gg}}  &= \frac{1}{\Delta \chi_{\text{g}}\chi_{\text{g}}^{2}}P_{\text{gg}}\left(\frac{\ell+0.5}{ \chi_{g}}, z_{g}\right) 
\end{aligned}
\end{equation}
where $n_{e0}$ represents the current mean number density of free electrons, $\chi_g$ denotes the assumed comoving side length of the observed box universe within a galaxy redshift bin, and $\Delta \chi_g$ is the comoving distance of each galaxy redshift bin, which represents the thickness of the redshift slice and is related to the redshift width $\Delta z_g$. $P_{\rm gg}, P_{\rm ee}, P_{\rm ge}$ represent the three-dimensional power spectra as functions of the magnitude $k$ of the three-dimensional Fourier wave number { k}, and we use halo models as described in \cite{Smith2018KSZTA} to calculate them. Notably, the electron profile $u_e(k|m,z)$ appearing in $P_{\rm ee}$ and $P_{\rm ge}$ is calculated using an active galactic nucleus (AGN) model \citep{2016JCAP...08..058B}, which accounts for the distribution of free electrons around galaxies influenced by AGN feedback. 

\begin{figure}
\centerline{\includegraphics[width=5in]{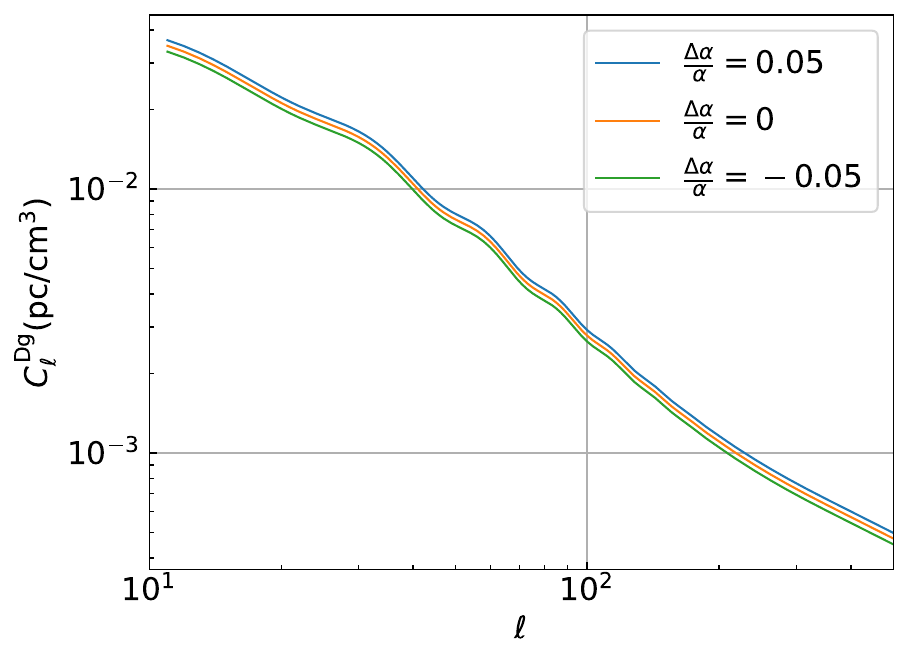}}
\caption{The simulated power spectrum of the DM-galaxy cross-correlation, \( C_{\ell}^\text{Dg} \) at $z=0.15$ with \(\text{N}_\text{FRB}=10^5\), is shown  with different values of \(\Delta \alpha / \alpha\). The blue line, positioned below the orange line, represents the case without the temporal evolution of \(\alpha\), while the blue and green lines correspond to evolution rates of \(0.05\) and \(-0.05\) for \(\alpha\), respectively.}
\label{fig:cl}
\end{figure}

In Fig. \ref{fig:cl}, we illustrate the impact of a non-zero variation in the fine-structure constant on the { simulated DM-galaxy cross-correlation angular power spectrum} at \(z_g=0.15\). As anticipated, a positive \(\Delta\alpha/\alpha\) increases the amplitude of the power spectrum (depicted by the blue line), whereas a negative \(\Delta\alpha/\alpha\) reduces its amplitude (shown by the green line). Accurate measurements of \(C_{\ell}^{\rm Dg}\) at small angular scales would substantially enhance the constraints on \(\Delta\alpha/\alpha\). 

In this study, we focus on the DM-galaxy angular power spectrum rather than the DM-DM power spectrum. The latter, being an integral over redshift, averages the effects of \(\Delta\alpha/\alpha\), thereby diluting its sensitivity and reducing the constraining power.

\section{Constraint results}\label{sec:3}

In this section, we present the constraints on the variation of the fine-structure constant with the cross-correlation angular power spectrum between foreground galaxy distributions and the DMs of background FRBs. Additionally, we explore the implications of these constraints on several cosmological phenomena, including the CDDR, the evolution of the CMB temperature, and the proton-to-electron mass ratio.

\subsection{The fine-structure constant}\label{sec:alpha}
We utilize the Markov Chain Monte Carlo (MCMC) method to constrain the time evolution of the fine-structure constant, \({\Delta \alpha}/{\alpha}\). The likelihood function employed in this analysis is given by:
\begin{equation}
\chi^{2} = \left(\hat{C}_{\ell}^{\mathrm{Dg,obs}} - C_{\ell}^{\mathrm{Dg,th}}\right) \Gamma^{-1}_{\ell,\ell^{\prime}} \left(\hat{C}_{\ell^{\prime}}^{\mathrm{Dg,obs}} - C_{\ell^{\prime}}^{\mathrm{Dg,th}}\right)^{\mathrm{T}},
\end{equation}
where the fiducial value is set to \(({\Delta \alpha}/{\alpha})^\text{fid}=0\), and the noise covariance matrix is \(\Gamma_{\ell,\ell^{\prime}} = \delta_{\ell, \ell^{\prime}} \left(N_{\ell}^{\mathrm{Dg}}\right)^{2}\). The noise power spectrum for \(C_{\ell}^\text{Dg}\) is expressed as:
\begin{equation}
\left({N}_{\ell}^\text{Dg}\right)^{2} = \frac{1}{(2\ell+1)f_{\rm sky}}\left[\left(C_{\ell}^\text{Dg}\right)^2+\left(C_{\ell}^\text{gg} + N_{\ell}^\text{gg}\right)\left(C_{\ell}^\text{DD} + N_{\ell}^\text{DD}\right)\right],
\end{equation}
where \(N_{\ell}^\text{gg} = 1/n^{2d}_g\) represents the noise power spectrum of galaxies. Here, we adopt parameters from a CSST-like spectroscopic survey with a galaxy number density of \(n_g = 3.4 \times 10^{-2} \, \text{Mpc}^{-3}\text{h}^{3}\) and a sky coverage fraction of \(f_\text{sky} \simeq 0.424\), consistent with the first redshift bin described in \cite{2019ApJ...883..203G}.

When observing the DM field, noise arises from discrete sampling and the variance in the measured DMs. This noise can be quantified using the noise power spectrum \( N_{\ell}^\text{DD} \), given by: \begin{equation}
N_{\ell}^\text{DD} = \frac{\sigma_\text{DM}^{2}}{n_{f}^{2d}}
\end{equation}
where \( n_{f}^{2d} \) represents the number density of FRBs per steradian, and \( \sigma_\text{DM}^2 \) denotes the total variance of the DMs. The total variance arises from three main sources: the Milky Way Galaxy, the host galaxy, and the IGM. The contribution from the Milky Way can be estimated using models such as NE2001 \citep{2002astro.ph..7156C} or YMW16 \citep{2017ApJ...835...29Y}, with its root mean square (RMS) commonly assumed to be \( 10 \, \text{cm}^{-3}\text{pc} \). The contribution from the host galaxy, encompassing its halo and interstellar medium, is subject to significant uncertainties. Meanwhile, the IGM contribution reflects the electron distribution along the intergalactic propagation path of the FRB.

{ The variations in \(\text{DM}_\text{IGM}\) and \(\text{DM}_\text{host}\) remain uncertain, and the precise determination of the DM RMS is not the primary focus of this study. Therefore, we adopt a statistical value for analysis with a DM variance of \( \sigma_\text{DM} = 300 \, \text{cm}^{-3}\text{pc} \) \citep{mcquinn2013locating,li2019cosmology}. Additionally, we assume a total of \( 10^5 \) fast radio bursts (FRBs). These assumptions are consistent with the simulation results provided in \cite{2015MNRAS.451.4277D}.}


For our analysis, we adopt \(\ell_\text{max}=500\) to account for the resolution limitations of the observed FRBs, and \(\ell_\text{min}=11\) to exclude large scales where the Limber approximation is no longer valid. The MCMC analysis is conducted using the \texttt{cobaya} package \citep{2021JCAP...05..057T} within the standard \(\Lambda\text{CDM}\) cosmological model. The parameter \({\Delta \alpha}/{\alpha}\) is explored over the range \([-0.1, 0.1]\) in this analysis.

\begin{figure}
\centerline{\includegraphics[width=5in]{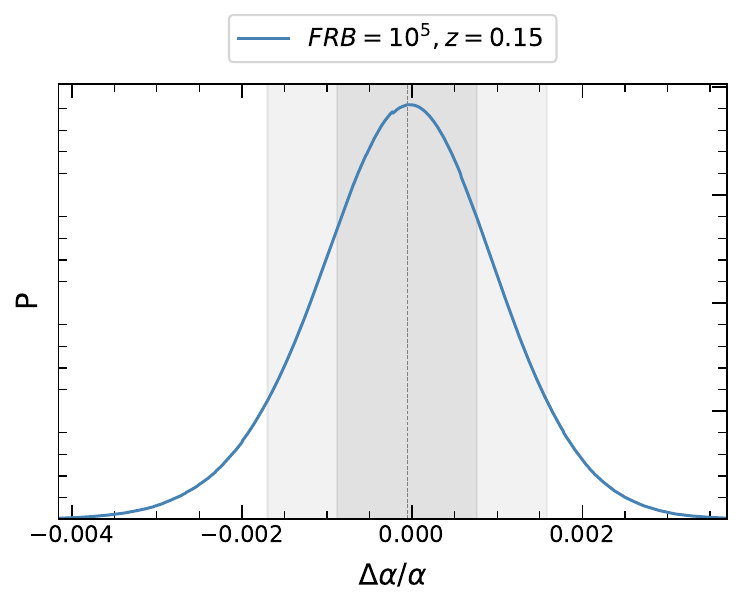}}
\caption{The 1D posterior distribution of \(\Delta \alpha / \alpha\) is obtained using the MCMC method at \(z_g = 0.15\), with a number of FRBs \(N_\text{FRB} = 10^5\) and a DM RMS of \(\sigma_\text{DM} = 300 \, \text{cm}^{-3}\text{pc}\). The grey line represents the best-fitting value of \(\Delta \alpha / \alpha\). The dark grey and light grey bands correspond to the \(1\sigma\) and \(2\sigma\) confidence levels, respectively.}
 \label{fig:mcmc}
\end{figure}

The results of our analysis are shown in Fig. \ref{fig:mcmc}, where the grey dashed line represents the best-fitting value for \({\Delta \alpha}/{\alpha}\). Our MCMC analysis yields a precise constraint on the time evolution of the fine-structure constant, with a standard deviation of \(\sigma({\Delta \alpha}/{\alpha}) = 0.0007\), at the effective redshift of the galaxy bin, \(z_g = 0.15\). Moreover, if the number of FRB sources is increased or the DM variance is reduced, the constraint on \(\Delta \alpha/\alpha\) can be improved further. { For example, when the number of FRBs reaches \(10^7\), the constraint on \(\Delta \alpha/\alpha\) is expected to improve by 40\% compared to the case with \(10^5\) FRBs.} This result underscores the robustness of our methodology and highlights the potential of using the DM-galaxy cross-correlation angular power spectrum to probe fundamental aspects of physics.

The temporal variation of the fine-structure constant, \(\alpha\), has been investigated in other studies using astronomical data. For instance, { \citet{WOS:000862992800009} analyzed} strong gravitational lensing and Type Ia supernovae observations, obtaining a constraint with a $1\sigma$ error of approximately $0.1$. Apparently, our work gives significantly tighter constraint on $\Delta\alpha/\alpha$. A more recent study by \citet{WOS:001251659300001} reported a variation in \(\alpha\) at the level of \(10^{-5}\), which utilized a dataset of approximately 110,000 galaxies with strong and narrow \([O\, \text{III}] \lambda \lambda 4959, 5007\) emission lines within a redshift range of [0, 0.95]. Their constraints slightly surpass the results of our study; however, with ongoing advancements in the precision of FRB observations, achieving tighter constraints on \(\alpha\) is expected to be feasible in the near future.

\subsection{Effects on CDDR and evolution of the CMB temperature}

We now examine the impact of the evolution of \(\alpha\) on other cosmological parameters, including the CDDR parameter \(\eta\) and the evolution parameter \(\beta\) of the CMB temperature.

It is commonly proposed that an extra scalar field \(\phi\) couples to other matter fields (such as the electromagnetic field), introducing a non-minimal multiplicative coupling in the action as follows: 
\begin{equation}
	S_{\mathrm{mat}}=\int d^{4} x \sqrt{-g} h_i(\phi)\mathcal{L}_{\mathrm{i}}\left(g_{\mu \nu}, \Psi\right),
\end{equation}
where \(h_i(\phi)\) characterizes the coupling between these two fields. General relativity is recovered when \(h(\phi) = 1\). Previous studies suggest that such couplings imply both the evolution of the fine-structure constant and photon number non-conservation along geodesics. The latter effect manifests as violations of the CDDR and deviations from the standard evolution of the CMB temperature \citep{1996PhRvD..54.2571L,10.1046/j.1365-8711.2000.03172.x,2014PhRvD..90l4064H}. 

As \(\alpha \sim h^{-1}(\phi)\) \citep{2002PhRvD..66d6007D,PhysRevLett.89.081601}, combining this relationship with Eq.~\ref{eq:alpha1}, the time evolution of \(\alpha\) can be expressed as:
\begin{equation}\label{eq:alpha2}
	\frac{\Delta \alpha}{\alpha}(z) \equiv \frac{\alpha(z) - \alpha_0}{\alpha_0}=\frac{h(\phi_0)}{h(\phi)}-1~.
\end{equation}

\begin{table*}
\centering
\tabcolsep=1cm
\renewcommand\arraystretch{1.4}
\caption{The best-fit values, along with the 68\% and 95\% confidence intervals, are provided. The analysis is conducted at a redshift of \(z = 0.15\). The parameters \(\eta\), \(\beta\), and \(\Delta \mu/\mu\) correspond to the violation of the CDDR, the deviation from the standard evolution of the CMB temperature, and the temporal variation of the proton-to-electron mass ratio, respectively. } 
\label{tab} 
\begin{tabular}{clc} \hline Parameters  &68\% limits & 95\% limits  \\ 
\hline $\eta$  &$1.0000\pm{0.0004}$& $1.0000\pm{0.0010}$ \\ 
$\beta$  &$0.0001\pm{0.0006}$& $0.0001\pm{0.0020}$ \\ 
${\Delta \mu}/{\mu}$  &$0.000\pm{0.002}$& $0.000\pm{0.006}$ \\ \hline 
\end{tabular}
\end{table*}

\subsubsection{Cosmic distance duality relation}

Next, we analyze the CDDR, which connects two fundamental geometric distances in observational cosmology: the luminosity distance (\(D_\text{L}\)) and the angular diameter distance (\(D_\text{A}\)). Under standard assumptions, namely photon number conservation along null geodesics and the validity of the reciprocity relation, these two distances are linked through the CDDR \citep{doi:10.1080/14786443309462220,2014PhRvD..90l4064H}:  
$D_\text{L}(z) = D_\text{A}(z)(1+z)^2$.

However, introducing the coupling term \(h(\phi)\) modifies the Maxwell equations to account for the interaction between a scalar field \(\phi\) and photons \citep{2014PhRvD..90b3017M}:  
\begin{equation}
\nabla_\nu\left(h(\phi)F^{\mu\nu}\right) = 0,  
\end{equation}
where \(F^{\mu\nu} = \partial^\mu A^\nu - \partial^\nu A^\mu\) is the Faraday tensor, and \(A^\mu\) is the four-potential. This modification ensures that photons still propagate along null geodesics, preserving the reciprocity relation \citep{PhysRevD.87.103530}, but it violates photon number conservation. Consequently, the observed flux \(F\) and source luminosity \(L\) are altered, leading to a modification in the luminosity distance, defined as \(D_\text{L} = \sqrt{L / 4 \pi F}\).  

By integrating the modified Maxwell equations in a flat Friedmann-Lemaître-Robertson-Walker (FLRW) spacetime, the expression for \(D_\text{L}\) is given by \citep{2014PhRvD..90b3017M}:  
\begin{equation}
D_\text{L}(z) = c(1+z)\sqrt{\frac{h(\phi_0)}{h(\phi(z))}}\int_0^z \frac{\mathrm{d}z'}{H(z')},  
\end{equation}
where the subscript \(0\) refers to the present epoch.  

In contrast, the angular diameter distance, \(D_\text{A}\), is derived via the standard null geodesic integral, unaffected by the coupling term, and retains its expression as in GR. As a result, deviations from the CDDR can be parameterized as \citep{2014PhRvD..90l4064H}:  
\begin{equation}
\frac{D_\text{L}(z)}{D_\text{A}(z)(1+z)^2} \equiv \eta(z) = \sqrt{\frac{h(\phi_0)}{h(\phi(z))}},  
\end{equation}
where \(\eta(z)\) quantifies the deviation from the CDDR and serves as a proxy for the scalar field’s time evolution.  

Constraints on \(\Delta \alpha / \alpha\) obtained from experimental measurements can thus be utilized to constrain \(\eta(z)\), and vice versa. Combining this relation with Eq.~\ref{eq:alpha2}, we establish a direct connection between \(\eta(z)\) and \(\Delta \alpha / \alpha(z)\):  
\begin{equation}\label{eq:alpha_eta}
	\frac{\Delta \alpha}{\alpha}(z)=\eta^2(z)-1~.
\end{equation}
From our MCMC analysis, we { obtain} a constraint on \(\eta\) at \(z_g = 0.15\), as presented in Table \ref{tab}. The standard deviation of \(\eta(z=0.15)\) is \(\sigma(\eta) = 0.0004\), representing a significant improvement compared to other prospective measurements of \(\eta\) in future experiments \citep{Matos_2024,Euclid:2020ojp,2019BAAS...51g..35R,LISA:2017pwj}. This result highlights the power of the DM-galaxy cross-correlation technique in providing precise constraints on deviations from the CDDR.

\subsubsection{The evolution of the CMB temperature}

In the framework of the standard hot Big Bang model, the evolution of the CMB temperature follows a linear relation with redshift, characterized by \(\beta = 0\):  
\begin{equation}\label{eq:tem1}
    T(z) = T_0 (1+z)^{1-\beta},
\end{equation}  
where \(T_0\) is the CMB temperature measured locally at \(z = 0\). A nonzero \(\beta\), however, suggests deviations from the standard model, such as non-adiabatic expansion or photon number non-conservation. These deviations also introduce a chemical potential \(\mu\) in the CMB spectrum, which quantifies departures from the ideal blackbody distribution.  

In this context, the inclusion of a coupling term \(h(\phi)\), representing photon number non-conservation, further modifies the propagation of CMB photons. This coupling leads to a departure from the blackbody spectrum, reflecting an out-of-equilibrium state of the CMB. The underlying mechanism is described by the Boltzmann equation, where the collision term explicitly incorporates the effect of \(h(\phi)\). A detailed derivation of this formalism is provided in \cite{2014PhRvD..90l4064H}.

By extending the photon number density \( n \), energy density \( \rho \), and temperature \( T \) to first-order perturbations, where \( x_i = x_{i,0} + \delta x_i \) (\( T, \rho, n \)), and \( x_{i,0} \) represents the unperturbed values when \( h(\phi) = 1 \), we can derive the temperature evolution:  
\begin{equation}\label{eq:tem2}
    \frac{\delta T}{T_{(0)}} = \frac{\frac{\delta\rho}{\rho_{(0)}} - \frac{540\zeta(3)^2}{\pi^6}\frac{\delta n}{n_{(0)}}}{4\left(1 - \frac{405\zeta(3)^2}{\pi^6}\right)},
\end{equation}  
where \( \zeta(3) \) is the Riemann zeta function evaluated at 3.  

At \( z_\text{CMB} \), where the CMB temperature is unaffected by the coupling, only zero-order terms exist. It can be shown that both \( \delta n / n_{(0)} \) and \( \delta \rho / \rho_{(0)} \) satisfy:  
\begin{equation}
    \frac{\delta n}{n_{(0)}} = \frac{\delta \rho}{\rho_{(0)}} = \delta h(\phi) \equiv \frac{h(\phi_{\text{CMB}})}{h(\phi)} - 1,
\end{equation}  
which implies from Eq. \ref{eq:tem2} that:  
\begin{equation}
    \frac{\delta T}{T_{(0)}} \simeq 0.12 \delta h(\phi).
\end{equation}  
Combining Eqs. \ref{eq:alpha2} and \ref{eq:alpha_eta}, we can relate \(\Delta \alpha / \alpha(z)\) to \(\delta h(\phi)\):  
\begin{equation}
    \delta h(\phi) = \frac{\eta^2(z)}{\eta^2(z_\text{CMB})} - 1 \simeq \frac{\Delta \alpha}{\alpha}(z) - \frac{\Delta \alpha}{\alpha}(z_\text{CMB}).
\end{equation}  
Thus, the CMB temperature evolution at redshift \( z \) becomes:  
\begin{equation}
    T(z) = T_{(0)} + \delta T = T_0 (1+z) \left[ 1 + 0.12 \frac{\Delta \alpha}{\alpha}(z) \right].
\end{equation}  
This equation establishes a direct and precise connection between the evolution of the fine-structure constant \(\alpha\) and deviations in the cosmic evolution of the CMB temperature. 

By comparing it with Eq. \ref{eq:tem1}, and assuming that the evolution of \(\alpha\) at \(z_\text{CMB}\) is negligible, we { obtain} a direct constraint on \(\beta\):  
\begin{equation}
    (1+z)^{-\beta} = 1 + 0.12 \frac{\Delta \alpha}{\alpha}(z),
\end{equation}  
where the constraint on \(\beta\) is determined directly from the observational limits on \(\Delta \alpha / \alpha\). We summarize the constraints on \(\beta\) obtained from the DM-galaxy cross-correlation angular power spectrum in Table \ref{tab}. At \(z_g = 0.15\), we achieve a standard deviation of \(\sigma(\beta) = 0.0006\), demonstrating the precision of our methodology. This constraint is significantly tighter compared to the results from experiments such as the Atacama Cosmology Telescope \citep{2021ApJ...922..136L}, which reported \(\beta = 0.017^{+0.029}_{-0.032}\). This highlights the potential of our approach in providing superior constraints on deviations from the standard adiabatic expansion model.

\subsection{Effect on proton-to-electron mass ratio}

Following \citet{coc2007coupled}, we consider a class of grand unification models where the weak scale is determined by dimensional transmutation. In these models, the variation of fundamental couplings is driven by a dilaton-type scalar field, resulting in a connection between the fine-structure constant \(\alpha\) and the proton-to-electron mass ratio \(\mu = m_p / m_e\). This relationship can be expressed as:  
\begin{equation}
\frac{\Delta \mu}{\mu} = \left[\frac45R-\frac3{10}(1+S)\right] \frac{\Delta \alpha}{\alpha},
\end{equation}
where \(R\) and \(S\) are a model-dependent proportionality constant that encapsulates the specifics of the unification scenario. For typical grand unification models, they are often estimated to be $(R,S)=(36,160)$ \citep{coc2007coupled}, which is consistent with the constraint from the atomic clocks measurement \citep{Juliao:2013xia}.  

Using the constraints on \(\Delta \alpha / \alpha\) derived from our analysis, we can translate these into constraints on the variation of \(\mu\), which provides another avenue to probe the underlying physical mechanisms and scalar fields influencing fundamental constants. The standard deviation of $\Delta\mu/\mu$ at $z=0.15$ is $\sigma(\Delta\mu/\mu)=0.002$. This tight constraint significantly enhances our ability to probe the underlying physical mechanisms and scalar fields that drive variations in fundamental constants, reinforcing the utility of the DM-galaxy cross-correlation as a sensitive probe of new physics.

\section{Conclusions}\label{sec:4}

The question of whether the fundamental constants of nature, such as the fine-structure constant \(\alpha\), evolve over time has been a long-standing issue in physics and cosmology. This study presented a cosmological test for the potential time evolution of \(\alpha\) using the cross-correlation between FRBs and a galaxy survey. Our analysis reveals that incorporating the evolution of the fine-structure constant introduces an additional factor of \(({\Delta \alpha}/{\alpha}(z) + 1)\) into the DM-related expressions. This modification provides a pathway to explore the impact of fundamental physics on cosmological observations and serves as a sensitive probe of the temporal evolution of \(\alpha\).

To isolate the contribution of electrons at the redshift of interest, we adopt a geometric model where galaxies act as the foreground within a narrow redshift bin $z_g=0.15$, while { mock FRBs} $z_f=0.4$ serve as a backlight beyond the galaxy. This configuration allows us to precisely target the electron density at the desired redshift. Using this framework, we derived the auto- and cross-power spectra for galaxies and the DMs, as well as accounted for the contribution of shot noise.  

In this letter, we performed a MCMC analysis to constrain the evolution of the fine-structure constant, \(\alpha\), characterized by \({\Delta \alpha}/{\alpha}\). Additionally, we imposed stringent constraints on related parameters, including the CDDR parameter \(\eta\), the evolution parameter \(\beta\) of the CMB temperature and the evolution of the proton-to-electron mass ratio. The main conclusions are as follows:
\begin{enumerate}
	\item Assuming the foreground galaxies are distributed within a thin redshift shell at \(z_g = 0.15\) with a number density of \(n_g = 3.4 \times 10^{-2} \, {\rm \text{Mpc}^{-3}\text{h}^{3}}\), { and a mock sample of background FRBs} is located at \(z_f = 0.4\) with a source number of \(N_{\rm FRB} = 10^5\) and a DM variance of \(\sigma_{\rm DM} = 300 \, {\rm pc/cm^3}\), we employ the DM-galaxy cross-correlation angular power spectrum to constrain the variation of the fine-structure constant, achieving a high precision of \(\sigma(\Delta\alpha/\alpha) = 0.0007\).  

\item Considering a non-minimal coupling between an extra scalar field and other matter fields, such as the electromagnetic field, photon number conservation along geodesics is violated. Consequently, the constraint on \(\Delta\alpha/\alpha\) can be translated into constraints on the violation of the CDDR and the CMB temperature-redshift relation. At \(z = 0.15\), we { obtain standard deviations for the parameters \(\eta\) and \(\beta\) as 0.0004 and 0.0006}, respectively, representing significantly tighter constraints compared to current observational results.  

\item Lastly, within the framework of grand unification models where the variation of fundamental couplings is driven by a dilaton-type scalar field, we establish a relationship between the temporal evolution of the proton-to-electron mass ratio and the fine-structure constant. This yields a tight constraint of \(\sigma(\Delta\mu/\mu) = 0.002\) at \(z = 0.15\).  
\end{enumerate}

These results highlight the utility of the DM-galaxy cross-correlation angular power spectrum in probing fundamental physics and cosmological principles. 

Moreover, the temporal evolution of fundamental constants has been the focus of extensive research due to its profound implications for both physics and cosmology. For instance, as first discussed in \citet{1999PhRvD..60b3515H}, variations in fundamental constants primarily impact the cosmic microwave background (CMB) through modifications to the recombination process. This has inspired numerous studies on the CMB temperature spectrum, including constraints on the electron mass \citep{2024PhRvD.110h3505B} and the chemical potential \(\mu\). Additionally, some studies suggest that uncertainties in the Hubble constant (\(H_0\)) could significantly affect constraints on the fine-structure constant \citep{WOS:001344562900001}. This highlights the interconnected nature of cosmological parameters and the importance of precise measurements. Investigating the temporal evolution of the fine-structure constant using diverse datasets \citep{WOS:000751345000002, WOS:000862992800009} offers a dual advantage: it provides deeper insights into the evolution of fundamental constants over cosmic time and serves as a robust test of alternative gravitational theories. This reinforces the need for cross-disciplinary approaches in exploring such foundational questions.

\begin{acknowledgments}
This work is supported by the National Natural Science Foundation of China, under grant Nos. 12473004 and 12021003, the National Key R\&D Program of China, No. 2020YFC2201603, the China Manned Space Program through its Space Application System, and the Fundamental Research Funds for the Central Universities.
\end{acknowledgments}

%





\bibliography{sample631}{}
\bibliographystyle{aasjournal}



\end{document}